\documentclass[prl,floatfix,twocolumn,showpacs,preprintnumbers,amsmath,amssymb]{revtex4}

\usepackage{graphicx,epsfig}
\usepackage{dcolumn}
\usepackage{bm}

\begin{document}

\preprint{APS/123-QED}

\title{Resonant X-ray diffraction studies on the charge ordering in magnetite}

\author{E.~Nazarenko$^{1,2}$, J.E.~Lorenzo$^1$, Y.~Joly$^1$, J.L.~Hodeau$^1$, D.~Mannix$^3$ and C.~Marin$^4$}
\affiliation{
$^1$Laboratoire~de~Cristallographie~CNRS,~BP~166X,F-38052~Grenoble~Cedex~09,~France \\
$^2$ Rostov~State~University,~Rostov-on-Don~344090,~Russia \\
$^3$XMaS~CRG,~European~Synchrotron~Radiation~Facility,~F-38043 Grenoble Cedex,~France. \\
$^4$DRFMC/SPSMS/Groupe~Mat\'eriaux,~CEA,~F-38054~Grenoble,~France.
 }

\date{\today}

\begin{abstract}
Here we show that the low temperature phase of magnetite is
associated with an effective, although fractional, ordering of the
charge. Evidence and a quantitative evaluation of the atomic
charges are achieved by using resonant x-ray diffraction (RXD)
experiments whose results are further analyzed with the help of ab
initio calculations of the scattering factors involved. By confirming the results obtained from X-ray crystallography we have shown that RXD is able to
probe quantitatively the electronic structure in very complex
oxides, whose importance covers a wide domain of applications.

\end{abstract}

\maketitle

Known in ancient times as lodestone and used to magnetize the
mariner's compass \cite{Hughes99}, magnetite (Fe$_3$O$_4$) is
still today the archetype compound of a number of physical
properties and applications. For instance, it is a promising
candidate for the development of highly sensitive
magneto-resistive devices in spin electronics. A sufficient
comprehension of magnetite is mandatory in view of its
applications and as a reference for similar effects in related
materials. Particularly, the magnetic and magneto-electric
properties of magnetite exhibit a conspicuous anomaly at T$_V$ =
121K that still defies understanding \cite{Walz02}. First
suggested by Verwey \cite{Verwey39} and expected in many oxides
\cite{Imada98,Coey04}, the low temperature phase transition in
magnetite has been associated to a charge disproportion on the
metal atoms sites although a direct confirmation has never been
evidenced in magnetite up to now. Theoretical predictions
\cite{Leonov04,Jeng04} have recently supported Verwey's scheme of
the localization of the charge (but on a more complex pattern) and
nuclear magnetic resonance \cite{Novak00} and M\"ossbauer
\cite{Berry98} experiments are compatible with different oxidation
states of the octahedral iron sites.  However direct confirmation
of charge disproportionation is still lacking.

\begin{figure}[!ht]
\begin{center}
\epsfig{file = 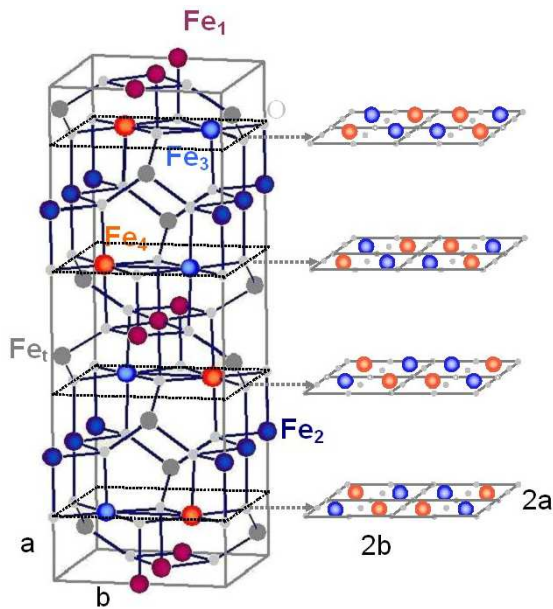, width=7.5cm }
\caption{\label{Structure} (Color online) (Left) A sketch of the
low temperature Fe$_3$O$_4$ structure in the orthorhombic space
group $Pmca$. There are six non equivalent iron atoms, two in the
tetrahedral sites (Fe$_t$) and four in the octahedral sites
(Fe$_1$, Fe$_2$, Fe$_3$, Fe$_4$) and it is understood that charge
ordering occurs at the octahedral sites. The actual structure is
known to be $Cc$ in a approximately $\sqrt{2}a_c \times
\sqrt{2}a_c \times 2a_c$ unit cell, with 16 unequivalent Fe-sites
at octahedral positions. However, no structural refinement in this
space group has been successful yet. (Right) Cell doubling along
$a$ ($\frac{1}{2}(a_c+b_c)$) and $b$ ($\frac{1}{2}(a_c-b_c)$) axis
of the $Pmca$ structure.  The color code tags represents Fe$_3$
and Fe$_4$ positions in the settings of the $Cc$ space group. This
CO model has been tested and greatly improves the quality of the
refinement.}
\end{center}
\end{figure}

From the structural point of view the determination of the atomic
positions issued from the metal-insulator transition challenges the
scientific community ever since Verwey's seminal work \cite{Verwey39,Walz02}. Progress in the
solution of the problem runs parallel to the development of new
and sophisticated experimental techniques as well as to the
implementation of refined computing codes and appropriate data
analysis strategies. Despite all these technical advances,
magnetite still remains a rather difficult case for conventional
crystallography: the symmetry lowering ($Fd\bar{3}m \rightarrow
Cc$, and $a_c \times a_c \times a_c \rightarrow \approx
\sqrt{2}a_c \times \sqrt{2}a_c \times 2a_c$, with $a_c$=8.394 \AA)
generates 8 and 16 non equivalent iron sites at tetrahedral and
octahedral positions, respectively, each one with its own atomic
charge. However, the final structure is not yet perfectly known.
The actually best refinement has been recently performed by Wright
and collaborators \cite{Wright01} (in space group $Pmca$, with
lattice parameters $\approx a_c/\sqrt{2} \times a_ c/\sqrt{2}a_c
\times 2a_c$). Complexity is greatly reduced in this structure
model (Fig. \ref{Structure}), where there are 6 non equivalent
iron atoms, two in tetrahedral sites (Fe$_t$) and four in
octahedral sites (Fe$_1$, Fe$_2$, Fe$_3$ and Fe$_4$). Only the
octahedral irons are supposed to undergo a charge ordering (CO).
Whereas Fe$_1$ and Fe$_2$ are at the centers of their respective
octahedra, Fe$_3$ and Fe$_4$, posses more distorted local
environments, the iron atoms are slightly off center. It can also
be noticed that these position shifts are responsible for the
doubling of the cell along the $c-$axis whereas the Fe$_1$ and
Fe$_2$ ordering has no effect on the cell doubled.

Resonant X-ray diffraction is a technique where both the power of
site selective diffraction and the power of local absorption
spectroscopy regarding atomic species are combined to the best
\cite{Hodeau01}. Reflections are recorded over some tens
electron-volts around the absorption edge of the element(s)
present in the material, where they show strong energy and angular
dependencies. This phenomenon is due to the virtual photon
absorption-emission associated with the resonant transition of an
electron from a core level to some intermediate state above the
Fermi level. By virtue of the dependence on the core level state
energy and the three dimensional electronic structure of the
intermediate state, this technique is specially suited to study
charge \cite{Murakami98,Joly03}, orbital
\cite{Zimmerman01,Wilkins03} or spin orderings \cite{Gibbs88} and
associated geometrical distortions
\cite{Templeton85,Dmitrienko83}. In the case of charge ordering,
we exploit the fact that atoms with closely related site
symmetries but with barely different charges exhibit resonances at
slightly different energies. Here we show that the sensitivity of
this effect allows for quantitative estimations of the charge
disproportion. Opposite to fluorescence or absorption
measurements, the power of diffraction relies on the capability of
detecting differences that are even smaller than the inverse
lifetime of the core hole level. Clearly, not all Bragg
reflections are sensitive to charge ordering, and the failure in
detecting CO signatures in previous RXD work on magnetite is
partly related to an incomplete choice of them
\cite{Garcia01,Subias04}.

Our RXD experiments were carried out at the U.K. CRG beamline
(XmaS) in the European Synchrotron Radiation Facility (Grenoble)
operating at photon energies between 7060 and 7220 eV. The X-ray
beam wavelength was selected by a Si(111) double crystal
monochromator with an energy resolution of 0.8 eV at the photon
energies of the experiment. After scattering an analyzer crystal
was used to filter out background coming from the sample (mostly
fluorescence). The two samples used have been prepared using the
vertical floating zone technique, with the help of an optical
furnace (Crystal System Inc. FZ-T-10000-H-VI-P-G). The Fe$_3$O$_4$
feed rods were prepared from high purity powder (Alfa Aesar
99,997\%), compacted into cylindrical shape under isostatic press
(up to 1200 Bar). 
The Verwey transition was characterized by SQUID
measurements and results have been given in ref. \cite{Delille05}.
The as-grown crystals exhibit a sharp bulk Verwey transition with
T$_V$ = 121 K. 
The scattering experiment was carried out in reflexion mode, off a
\{110\} surface for both samples and at 50K.

\begin{figure}[!ht]
\begin{center}
\epsfig{file = 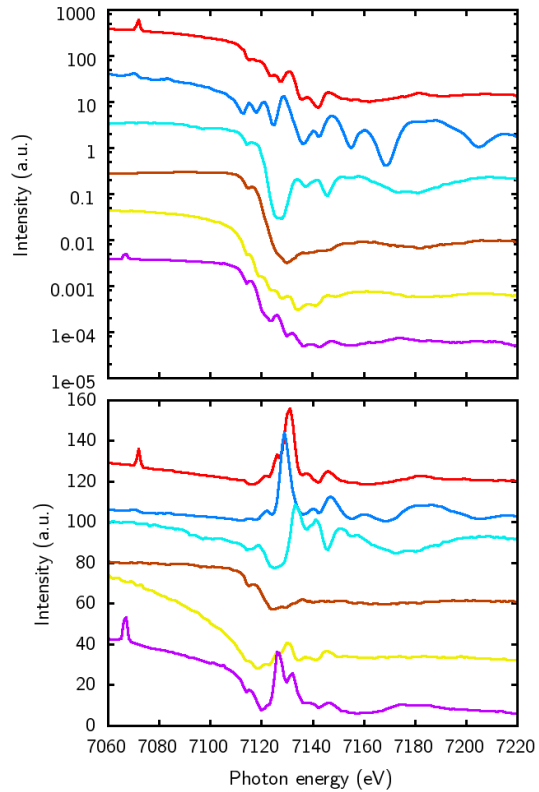, width=7.cm }
\caption{\label{Absorption} (Color online) (Top) Energy variation
of the intensities at the peak for a number of Bragg reflections
around the Fe K-edge and at T=50K. From top to bottom the
reflections are: (-4~4~1), (-4~4~2), (-4~4~3), (-4~4~4), (-4~4~5)
and (-3~3~3/2). Reflections are indexed according to the Miller
indices of the high temperature cubic phase. Data are completely
reproducible between the two studied samples. The small peaks at
around 7070 eV in the scans of the (-4~4~1) and (-3~3~1.5)
reflections are multiple scattering events that can be shifted in
energy by a slight change in the azimuth angle. The log scale
allows to appreciate the quality of the data, mainly above the
edge. (Bottom) The same reflections corrected from absorption.
Actual intensities have been normalized so as to fit into the
graph.}
\end{center}
\end{figure}

RXD experiments must be performed on a single crystal, if a good
signal-to-noise ratio is required. Traditionally one of the major
difficulties of crystallographic work on magnetite has been to
overcome a very strong self absorption, twinning, extinction and
the presence of multiple-scattering. These issues can be easier
overcome in RXD studies than in conventional single crystal
diffraction experiments. Indeed, parameter refinement in the
former is carried through the analysis of the energy dependence of
the photons scattered within each of the measured Bragg
reflections, the absolute scale being a unrefined proportionality
constant. In order to minimize the number of crystallographic
domains, a magnetic field of 0.3T has been applied parallel to one
of the [001] directions that uniquely defines the $c-$direction.
Multiple-scattering events are eliminated by the measurement of
the energy spectra at different azimuths and crystal settings. Raw
data have been plotted in figure \ref{Absorption}, top. Finally,
and more important, the absorption correction (figure
\ref{Absorption}, bottom) is carefully performed by using the
energy dependence of the intensities of several very strong Bragg
reflections as a measure of absorption. The anomalous contribution
of these reflections is relatively small and can be easily
computed. 

To account for the uncertainty of the crystallographic structure
and the fact that the charge ordering must be disentangled from
the associated atomic displacements, a complete methodology was
developed. It needs a very important set of experimental data,
first principle simulations and the use of objective confidence
factors for comparing experiment and theory  \cite{philip,horsky}.
Thus, no less than 50 reflections, were recorded over an energy
range of 120 eV around the Fe K-edge and at T=50K. This collection of peaks is at the variance of the RXD data collected in refs. \cite{Garcia01,Subias04} and use to dismiss CO in Fe$_3$O$_4$. 32 reflections
were kept for the optimization of the parameters, 12 were
reflections indexed in the $Cc$ cell (not used in this refinement)
and 6 strong reflections were used for normalization purposes. The
calculations are performed using the state of the art code FDMNES
\cite{Joly01,code}. It calculates the final states and the
resulting structure factors with first principal theories. In the
present study, the multiple scattering theory was used. This code
has already been efficient in the simulation of different spin,
orbital, charge or geometrical ordering phenomena
\cite{Joly04,Benfatto99}.

\begin{figure}[!ht]
\begin{center}
\epsfig{file = 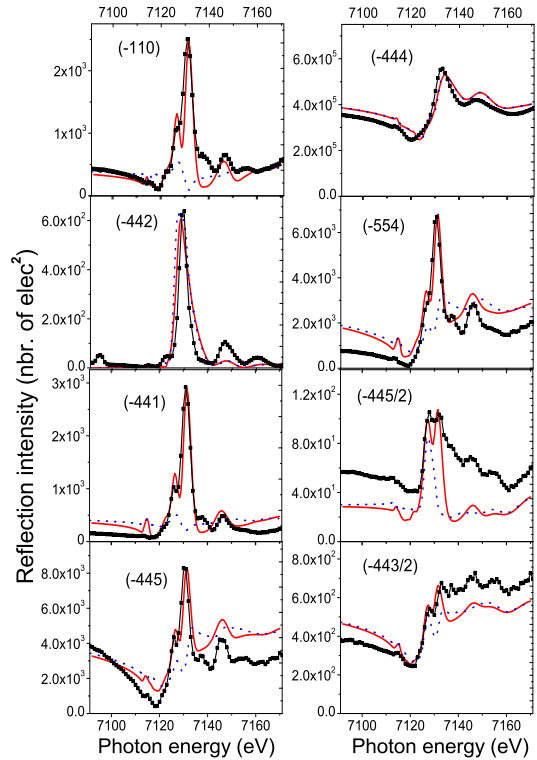, width=8.5cm }
\caption{\label{Refinement} (Color online) Some of the
experimental and calculated resonant X-ray diffraction peak
spectra in Fe$_3$O$_4$ at T=50K. Experimental data have been
corrected for absorption. Shown are the experimental data (black
dots), the calculated spectra with (red line) and without (blue
dots) charge ordering. Some reflections such as the (-1~1~0) and
the (-4~4~1) clearly display the \textit{derivative effect} (see
text) and are therefore very sensitive to the CO. Some others, as
the (-4~4~2), exhibit a line shape that does not arise from the
charge ordering. It is an extinct Bragg reflection of cubic phase
that displays features related to the anisotropy of the Fe-form
factors\cite{Templeton85,Dmitrienko83}. Strong reflections, such
as the (-4~4~4), were used for normalization purposes.
Half-integer reflections, such as (-4~4~5/2), are sensitive to the
charge difference between Fe$_3$ and Fe$_4$. Reflections are
indexed according to the high temperature cubic phase settings.}
\end{center}
\end{figure}

Trial calculations were performed for the two low temperature
phase structures proposed by Iizumi \textit{et~al.}
\cite{Iizumi82}. As expected, calculations in the non
centrosymmetric space group ($Pmc2_1$) produced a modest agreement
with the experimental data (not shown). Calculations in the $Pmca$
structure produces a far better agreement and in the following we
have used the recent results of ref. \cite{Wright01} with the
charge occupancies on the $3d$ iron sites as the only adjustable
parameters. From the local structures we conjecture that charge
ordering can be characterized by two different charge
disproportions, $\delta_{12}$ and $\delta_{34}$, between the
centered Fe$_1$ (charge state represented as
Fe$^{+(2.5-\delta_{12})}$) and Fe$_2$ (Fe$^{+(2.5+\delta_{12})}$)
sites and the off-centered Fe$_3$ (Fe$^{+(2.5-\delta_{34})}$) and
Fe$_4$ (Fe$^{+(2.5+\delta_{34})}$) sites, respectively. Tests on
other orderings, including the tetrahedrally-iron coordinated or
other combinations of octahedral iron atoms, have systematically
yielded rather poor confidence facto rs.

\begin{figure}[!ht]
\begin{center}
\epsfig{file = 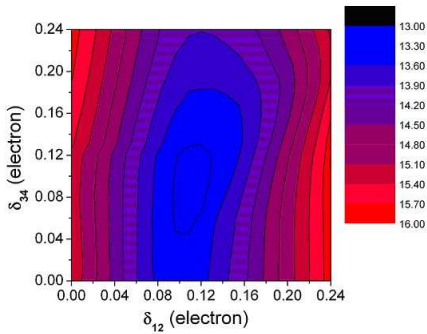, width=6.cm }
\caption{\label{Result} (Color online) Iso-value of the confidence
factor, the metric D$_1$ of Philip and Rundgren \cite{philip}. The
best agreement is obtained for $\delta_{12}$ = 0.12$\pm$0.025
electrons and $\delta_{34}$ = 0.10$\pm$0.06 electrons. It
corresponds to the 5.38, 5.62 and 5.40, 5.60 charge occupancies of
Fe$_1$ to Fe$_4$, respectively.}
\end{center}
\end{figure}

The most important result is that a significant improvement of the data fitting is achieved when an explicit CO is introduced
(Fig. \ref{Refinement}). Reflections such as (-4~4~1) are highly
sensitive to the magnitude of $\delta_{12}$. All of them exhibit a
double structure around 7126-7131 eV which can be taken as a
signature of the charge ordering phenomenon as seen by RXD. They
are a clear example of the \textit{derivative~line~shape},
characteristic of CO as likewise observed in other compounds
\cite{Joly03,Zimmerman01}. Indeed, for a specific charge pattern
and the reflections under consideration, the partial structure
factor associated with oppositely charged iron atoms almost
cancel. The optimization of the parameters at the Fe$_1$-Fe$_2$
sites gives a $\delta_{12}$ = 0.12$\pm$0.025 electrons which
corresponds to a shift of the absorption edge by 0.9 eV. This
result is particularly robust and remains unchanged whether a
large or a small data basis is used in the refinement. The effect
of $\delta_{34}$ is contained in half-integer reflections and even
more in reflections indexed in the $Cc$ cell. Interestingly, and
despite of the lack of knowledge about the actual lattice
distortion yielding a $Cc$ unit cell, the charge can be refined in
this space group (see Fig. \ref{Structure}, right, for the charge
distribution compatible with the $Cc$ symmetry) by using
reflections indexed for the $Pmca$ symmetry. The value of
$\delta_{34}$ is associated with a large uncertainty:
0.10$\pm$0.06 electrons, most probably reflecting the fact that
each of the Fe$_3$ and Fe$_4$ positions gives rise to four
nonequivalent sites, being not considered in this work. To obtain
a precise estimate of $\delta_{34}$ would require a concomitant
knowledge of the atomic displacements producing the $Cc$ structure
and the use of reflections indexed in the $\sqrt{2}a_c \times
\sqrt{2}a_c \times 2a_c$ unit cell. Figure \ref{Result} represents
the confidence factor plotted in the form of a contour plot. All
the refined charges are in very good agreement with the
theoretical predictions using LDA+U calculations
\cite{Leonov04,Jeng04}. The contours run nearly parallel to the
$\delta_{34}$ axis meaning that there is little correlation
between both charge orderings.

In this work we have solved the long standing controversy of
whether or not actual charge ordering develops at the octahedral
Fe-sites of Fe$_3$O$_4$ as a result of the metal-to-insulator
(Verwey) phase transition. Several conclusions can be drawn out.
First, we have shown that significant, non-integer charge
localization occurs at the metal sites in the low temperature
phase of magnetite. This result is robust and independent of the
uncertainties of the actual crystallographic structure. In view of
the short interaction time of the photons, $\approx$ 10$^{-16}$
s., we conclude that the ordering is purely static. Second, the
results of our experiment and additional calculations considering
the magnitudes of the charge are in good overall agreement with
those resulting from the bond valence sum method (BVS). Third, CO
at Fe$_1$ and Fe$_2$ sites gives rise to a [001]$_c$ charge
modulation whereas CO at Fe$_3$ and Fe$_4$ originate the
[00$\frac{1}{2}$]$_c$ ordering, and both are relatively uncoupled.
Fourth, the magnitude of the charge is in agreement with the
several band structure calculations \cite{Leonov04,Jeng04} which
implies that the phase transition, at least the ordered part, can
be explained under electronic considerations alone.

The authors want to thank V. Dmitrienko and S. Fratini for helpful
discussions and A. Kirfel for a critical reading of the manuscript.

\textit{Note Added.-} While the mansucript was with referees we became aware of a RXD  paper \cite{Goff05} where evidence for the CO has been found, thus supporting our results.

\end{document}